\documentclass[useAMS,usenatbib]{mn2e}
\usepackage{txfonts,graphicx,amssymb,rotating,epsfig,epstopdf}

\def\msun{{\,{\rm M}_\odot}}
\def\kms{km$\,$s$^{-1}$}

\def\simlt{\lower.5ex\hbox{$\; \buildrel < \over \sim \;$}}
\def\simgt{\lower.5ex\hbox{$\; \buildrel > \over \sim \;$}}

\title[Dynamics rebut cold-disc origin of GC stars]{Stellar dynamical evidence against a cold disc origin for stars in the Galactic Centre}

\author[J.\ Cuadra et al]{Jorge Cuadra$^{1}$\thanks{e-mail: 
jcuadra@jilau1.colorado.edu}, Philip J.\ Armitage$^{1,2}$, Richard D.\ Alexander$^{3,1}$
\\
$^{1}$JILA, Campus Box 440, University of Colorado, Boulder, CO 80309, USA \\
$^{2}$Department of Astrophysical and Planetary Sciences, University of Colorado, Boulder CO 80309, USA \\
$^{3}$Leiden Observatory, Universiteit Leiden, Niels Bohrweg 2, 2300 RA, Leiden, the Netherlands}

\begin{document}

\date{Accepted XXX. Received XXX; in original form XXX}

\pagerange{\pageref{firstpage}--\pageref{lastpage}} \pubyear{2008}

\maketitle

\label{firstpage}

\begin{abstract}
Observations of massive stars within the central parsec of the Galaxy
show that, while most stars orbit within a well-defined disc, a
significant fraction have large eccentricities and / or inclinations
with respect to the disc plane. Here, we investigate whether this
dynamically hot component could have arisen via scattering from an
initially cold disc -- the expected initial condition if the stars
formed from the fragmentation of an accretion disc.  Using N-body
methods, we evolve a variety of flat, cold, stellar systems, and study
the effects of initial disc eccentricity, primordial binaries, very
massive stars and intermediate mass black holes. We find, consistent
with previous results, that a circular disc does not become eccentric
enough unless there is a significant population of undetected
100--1000~$\msun$ objects. However, since fragmentation of an
eccentric disc can readily yield eccentric stellar orbits, the
strongest constraints come from inclinations.  We show that {\em none}
of our initial conditions yield the observed large inclinations,
regardless of the initial disc eccentricity or the presence of massive
objects. These results imply that the orbits of the young massive
stars in the Galactic Centre are largely primordial, and that the
stars are unlikely to have formed as a dynamically cold disc.
\end{abstract}

\begin{keywords}
{ stellar dynamics -- Galaxy: centre -- methods: N-body simulations}
\end{keywords}

\section{Introduction}
Observations of the Galactic Centre (GC) have revealed a population of young ($\sim$5~Myr) 
massive stars orbiting within a fraction of a parsec of the $M_{\rm BH} \approx 3 \times 10^6 \msun$ supermassive black hole Sgr~A*. 
The existence of any young stars so close to the black hole is a puzzle, since the tidal 
field would prevent star formation from gas clouds with densities $\rho < M_{\rm BH} / 
[(4 \pi / 3) r^3]$. This critical density is $\sim 10^{10}\,$cm$^{-3}$ at $0.1\,$pc, far higher 
than the densities of normal molecular clouds. The dynamics of the stars are also 
unusual and interesting. Although the innermost `S-stars' appear to be consistent 
with isotropic orbits, many of the young stars at a distance of the order of $0.1\,$pc 
lie on a single plane \citep{Levin03,Genzel03a,Paumard06,Lu06} and rotate 
clockwise on the sky. The eccentricities of stars in this 
clockwise disc are significant. The average value is $e \approx 0.35$, with the eccentricities
of several stars going above 0.5 \citep{Beloborodov06, Lu06}.
Moreover, 
the stars that do {\em not} belong to the clockwise disc (which, by definition, have 
large inclinations that in many cases exceed 90$^\circ$) have even larger 
eccentricities, perhaps as large as $e \approx 0.8$ for stars on counter-clockwise 
orbits \citep{Paumard06}\footnote{Note that there is some dispute as to whether 
there is evidence for a second, counter-clockwise 
disc of stars. This question is not important for our purposes. We rely on 
observations only insofar as they demonstrate that there is (at least) one disc, 
and that some stars have orbits that do not coincide with any suggested disc 
plane. These aspects are uncontested.}.

Two classes of theories have been advanced to explain the origin of the disc 
stars. In the first, the stars form {\em in situ} from a massive accretion disc 
in which the density is large enough to exceed the tidal limit \citep{Levin03, Milosavljevic04, NC05,
Paumard06}. Accretion discs at these radii around 
black holes are known to be vulnerable to fragmentation and star formation provided only 
that they can cool on a dynamical timescale \citep{Paczynski78,Kolykhalov80,
Shlosman89,Gammie01,Rice03,Goodman03}. The second model posits that the stars formed 
outside the GC region but migrated in due to dynamical friction 
\citep[e.g.][]{Gerhard01,Kim03,Kim04,Gurkan05,Berukoff06}. Understanding the predicted 
dynamics of these models in more detail may assist in discriminating between them.

Fragmentation of a thin accretion disc yields, almost by construction, starts 
with the basic disc-like dynamics observed in the GC. 
Within such a model those stellar orbits that do not belong to 
any well-defined disc have to be 
generated after the disc fragments, perhaps via scattering.  There is, however,
only just enough time for significant dynamical evolution to occur. 
\cite{Alexander07} found that even with a top-heavy mass function it was 
hard to attain the large eccentricities observed. This problem can be 
circumvented if the disc fragments while it is still 
eccentric \citep{NCS07,AACB08}, but it is still difficult to explain the 
high inclinations. Here, we investigate 
whether stellar dynamical evolution from {\em any} plausible cold disc initial 
conditions (i.e. those for which the local velocity dispersion is small 
compared to the orbital speed) can explain the observed stars. We consider 
both circular and eccentric initial conditions, and allow for the 
possibility of a high primordial binary population and additional 
scattering from unobserved massive objects (short-lived massive stars, 
or intermediate mass black holes).

\section{Numerical Method}
\label{sec:method}

We model the stellar dynamics of a disc of stars using the {\sc mercury6} code, 
which is optimized to study systems dominated by a central mass \citep{Chambers99}. 
In order to treat close encounters reliably we adopt the optional Bulirsch--Stoer integrator, 
in its version for conservative systems.

We simulate stellar discs extending radially from 0.05 to 0.15~pc.  
\footnote{Young stars are observed further away, but their density decreases rapidly and their scattering time-scales become longer with radius, so we do not expect the omission of stars at $r > 0.15\,$pc to alter our results. }
The discs are
composed of 100 stars randomly distributed following a surface density profile $\Sigma(r)
\propto r^{-1}$, roughly as observed in this radial range, with the mass of each star equal to $25\msun = 8.3 \times10^{-6} M_{\rm BH}$. In most of our models the stars are initially rotating in circular
Keplerian co-planar orbits around the central object, with small random perturbations such that the
thickness of the disc is $h/r = 0.01$.  The exact value of this perturbation is not
important, as the evolution of the disc is initially rapid.  We integrate the
systems for a time roughly equivalent to 2700 orbits at $r = 0.1\,$pc, almost 5~Myr. 
Since $N$ is rather small, 
we run three simulations with each set-up to monitor the variance in the outcome.

Our model system includes the number and masses of the 
{\em observed} population of stars in the GC. Although observational evidence 
favours a top-heavy mass function \citep{NS05,Paumard06} there could still be 
undetected lower-mass stars in the same region. Low mass stars, however, always act to 
{\em damp} the dispersion of the observed massive stars \citep{Alexander07}, so 
in ignoring them our models yield an upper limit to the rms eccentricities 
and inclinations that result from scattering.

\section{Simulations}

\subsection{Single stars}
\label{sec:singles}

\begin{figure}
   \centerline{\epsfig{file=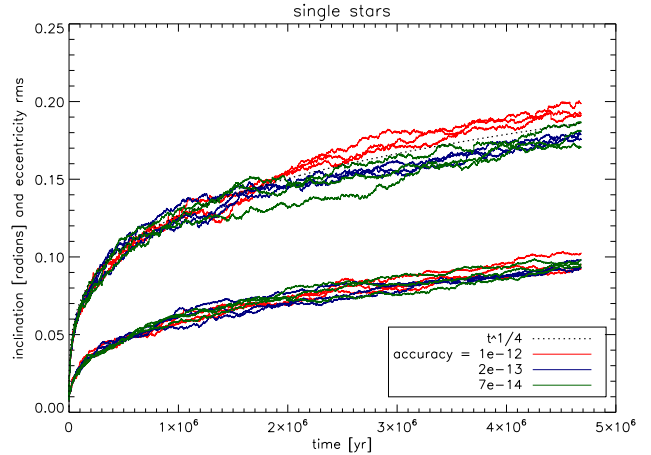,width=.49\textwidth}} 
   \caption{Eccentricity (upper curves) and inclination (lower curves) rms values from the nine realisations of discs of single
   stars. There is a small bias toward larger dispersion in the runs with less
   stringent accuracy requirements. For comparison, a curve $e \propto t^{1/4}$ is also shown.}
   \label{fig:singles}
\end{figure}

To establish a baseline for subsequent comparisons, we first model the evolution 
of a disc of single stars similar to that considered previously \citep{Alexander07}. 
We also use this simple system to test the numerical convergence of the integrator, 
by running three sets of models with different integrator tolerances. Each set 
includes three independent realisations of the initial conditions. Figure~\ref{fig:singles} shows the evolution of the rms values of the 
eccentricity and inclination for each simulation. We find that convergence 
is obtained for an accuracy parameter specified by {\tt accuracy}~$=2 \times 10^{-13}$, 
and that results accurate to the 10\% level, adequate for our purposes, can be 
obtained with {\tt accuracy}~$=10^{-12}$. We use the latter value for most of 
our subsequent runs.

As expected theoretically, the eccentricity and inclination rms grow roughly as $t^{1/4}$, with
a proportionality $e_{\rm rms} \approx 2 i_{\rm rms}$ \citep[e.g.,][]{Lissauer93}, where $i$ is measured in radians. To compare quantitatively with 
previous work, we fit the average eccentricity rms from all nine
simulations with the analytical solution, 
\begin{equation}
\frac{d\sigma_*}{dt} = \frac{G^2 N_*M_*^2 \ln \Lambda_*}{C_1 r \Delta r t_{\rm orb} \sigma_*^3},
\end{equation}
where $\sigma_*$ is the velocity dispersion of a stellar disc of $N_*$ stars each of mass $M_*$,
orbiting at radius $r$ with radial extension $\Delta r$.  The quantity $\ln \Lambda_*$ is the Coulomb
logarithm, $t_{\rm orb}$ the orbital period at $r$, and $C_1$ a geometrical constant of order
unity \citep[e.g.,][]{Alexander07}.  We find a value of $C_1 = 1.65$, in good agreement with the
1.85 that \cite{Alexander07} found from their high resolution runs.  This is a very small
difference in practice, as the eccentricity rms value scales only as $C_1^{-1/4}$. In 
agreement with prior work, we conclude that an initially circular disc set up to match 
the numbers and masses of observed stars in the GC attains an rms eccentricity of around 
0.2 within a few million years. This is too small to explain the larger eccentrities 
observed in the GC population.

\subsection{Primordial binaries}
\label{sec:binaries}

\begin{figure}
   \centerline{\epsfig{file=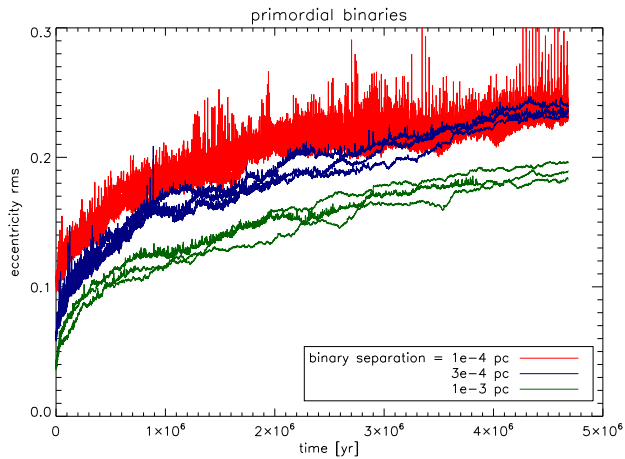,width=.49\textwidth}} 
   \caption{Eccentricity rms values from the nine different simulations of discs of binary stars.  The tighter the binaries are initially, the more dispersion the system acquires.}
   \label{fig:binaries}
\end{figure}

A small number of binary stars have been identified in the Galactic Centre 
\citep{Martins06,Peeples07,Rafelski07}, and, given our current knowledge of disc fragmentation, 
it is possible that the primordial binary fraction resulting from a fragmenting 
disc could be significant. Encounters between stars and hard binaries tend 
to shrink the binary orbit, releasing energy that can heat the system 
\citep{Heggie75,Perets07}\footnote{Supernovae can also heat the disc, as the surviving companion keeps the velocity it had in the binary.  \cite{Perets07}, however, showed that even with  $100\,$\kms kicks no significant inclinations are reached.}. To study whether binary heating could be significant in the GC we consider an idealised model in which the initial binary fraction is 
100\%. We replace the disc of 100 single stars with one made up of 50 equal mass 
binaries, whose centres of mass are initially in circular orbits about the black 
hole. For the fiducial case the binary orientations lie in the disc plane. In each simulation all binaries have the same initial value of semi-major axis, which we 
vary between 10$^{-3}$~pc and 10$^{-4}$~pc.  A binary with semi-major axis $10^{-4}\,$pc is hard (and hence will shrink and give energy back to the system) if the stellar velocity dispersion $\sigma_* \simlt 20\,$\kms, which corresponds to $e \sim 0.1$.  For numerical 
reasons we are only able to track binaries as they shrink down to a minimum size 
which we set at $3 \times 10^{-6}$~pc. If any stars get closer than this 
limit, the code merges them into a single object conserving energy and 
momentum. 

Figure~\ref{fig:binaries} shows the evolution of the rms value of the eccentricity for the 
binary runs
\footnote{ The average here is
defined not over stars, but rather over `objects' that could be either bound binaries or single
stars.  It is very hard to determine whether the
observed stars are actually singles or binaries, so the value we are plotting is what can be
compared to current observations.}.
 When the binaries' initial separations are $10^{-3}\,$pc they
are only marginally bound and most of them quickly get disrupted. As a result the system behaves
as if it were made up of single stars from the start, and the resulting eccentricities and 
inclinations are essentially identical to those discussed in the preceeding section. 
For initially tighter binaries, we see a trend where closer binaries produce more 
eccentric orbits. The plots also get noisier, as the instantaneous Keplerian orbital elements of a few stars are strongly perturbed during close encounters.
The changes are relatively modest, and even a disc made up entirely of the 
tightest binaries yields a final eccentricity that is only $\sim 25$\% larger than 
the single star case. The rms value of the inclination
is again roughly half of the eccentricity. Initial binary orientations that are 
not aligned with the disc plane yield modestly smaller inclinations and 
eccentricities than the aligned case. 
We interpret this as implying a reduced cross-section for binary--binary encounters in the non-aligned case.  
We also experimented with decreasing the 
minimum distance below which stars are assumed to merge. A reduction in this 
scale by a factor of three resulted in slightly smaller eccentricities, but 
we were unable to demonstrate convergence by evolving systems with even smaller 
merger radii due to numerical 
difficulties. Accordingly, we do not attach physical significance to this difference.  In reality, the dynamics of any binaries that shrink to such small 
scales (which are less than an AU) would likely be affected by the presence of 
circumstellar discs left over from their formation. We cannot model such 
effects, but it is hard to see how they could increase the overall stellar disc 
eccentricity.

\subsection{Eccentric flat discs}
\label{sec:eccentric}

\begin{figure}
   \centerline{\epsfig{file=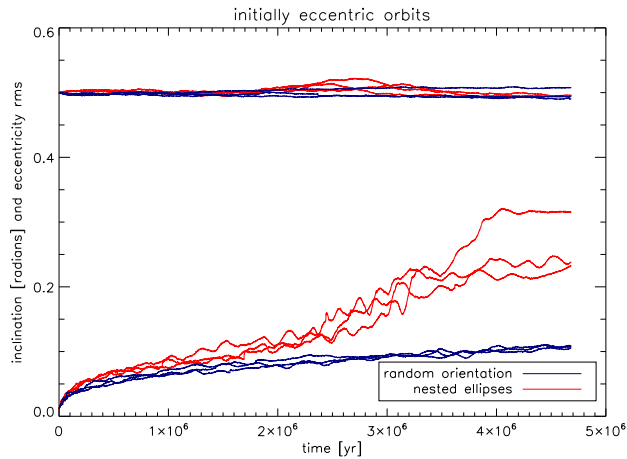,width=.49\textwidth}} 
   \caption{Eccentricity (upper curves) and inclination (lower curves) rms values from the eccentric disc runs. The eccentricity rms is roughly constant and the inclinations remain small in the more realistic case (see Section~\ref{sec:eccentric}).}
   \label{fig:ecc}
\end{figure}

If the stars originated from the fragmentation of a self-gravitating disc, they did not
necessarily form after the disc became circular. It is possible that gas settled in a plane
around the black hole and fragmented before it had time to circularise.  In
this case, the stars would initially have inherited the eccentricity of the gas 
\citep{NCS07, AACB08}.   

To study the growth of inclinations in this scenario, we ran several simulations 
with initially eccentric stellar discs. The stars were set up in orbits with $e = 0.5$, 
with a surface density $\Sigma(a) \propto a^{-1}$ between 0.05 and 0.15~pc. We considered 
both the case where the angle of pericentre of the eccentric orbits was initially constant 
(so all the ellipses are nested), and the case where this angle was randomly distributed 
(in this case the in-plane velocity dispersion is initially large). 
The aligned case is the expected outcome immediately after the fragmentation of an eccentric gaseous disc, but the $M_{\rm cusp} \sim 10^5 \msun$ worth of old stars between 
0.05--0.15 pc \citep{Genzel03a} will lead to differential precession and randomly oriented 
orbits on a time-scale $\sim (M_{\rm BH}/M_{\rm cusp})t_{\rm orb}$.  This time-scale is very short compared to the current age of the stellar system, so the more realistic set up for our simulations {\em without} the cusp potential are those with randomly chosen pericentre angles.

Figure~\ref{fig:ecc} shows the results for these simulations, with the blue curves depicting the 
case of randomly aligned orbits. The typical eccentricity remains fixed at $e \approx 0.5$, while the inclination reaches a typical value of 0.1, just as in the circular case (Sect.~\ref{sec:singles}). The runs starting with aligned orbits (red curves) have a larger spread in
eccentricities (although still $e_{\rm rms} \approx 0.5$), with approximately 10 of the innermost stars acquiring large eccentricities and inclinations as a result of the interaction with the eccentric potential of the rest of the stars.  This seems to be a resonant relaxation effect \citep{Rauch96}, a mechanism already invoked by \cite{Hopman06} and \cite{Levin07} to explain the high eccentricities and inclinations of the innermost `S-stars'.  The fraction of the stars that get high $i$ orbits is still too small to match the GC population, and, as explained above, we expect the background stellar potential to substantially weaken this resonant process.

\subsection{Influence of Massive Stars}
\label{sec:massive}

\begin{figure}
   \centerline{\epsfig{file=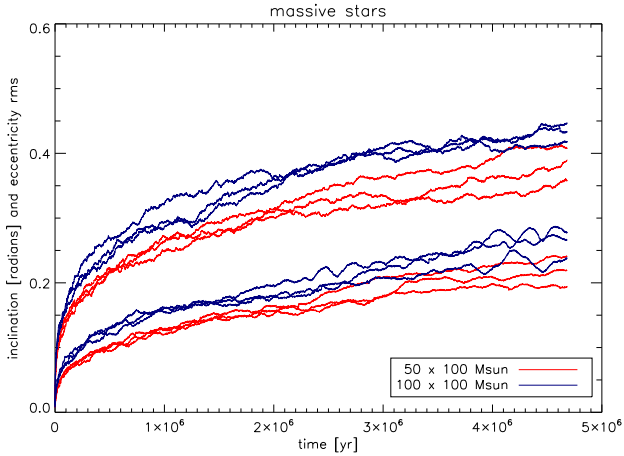,width=.49\textwidth}} 
   \caption{Eccentricity (upper curves) and inclination (lower curves) rms values from the six different simulations of discs of stars that included a significant number of very massive stars.  The very massive stars, however, are not included in this analysis, as they are not observed and are supposed to have already died.  The typical eccentricity reaches an interesting value, but the inclinations are still too low to match the GC observations.}
   \label{fig:massive}
\end{figure}

So far we have concentrated on the stars that are currently observed.  However, it is possible that many more massive stars were born but have already exploded as supernovae.  To take into account this effect, we ran a set of simulations where -- in addition to the 100 25-$\msun$ stars previously considered -- we include an extra number of
100~$\msun$ stars.  These stars are included throughout the simulation but are not taken into account in the final analysis,
so in practice it is as if these very massive stars disappeared a very short time ago.  As shown by \cite{Alexander07}, the
more massive stars will sink toward the mid-plane, while the lower mass stars will acquire higher eccentricities and
inclinations.

We ran several simulations with either 50 or 100 of the 100~$\msun$ stars.  Figure~\ref{fig:massive} shows the eccentricity and inclination rms {\em for the 25$\msun$ stars}. The eccentricities now reach a value $e \approx 0.4$, comparable to that observed.  The inclinations, however, are still too low to explain the non-disc population in the Galactic centre.  There is not much difference between the cases with 50 and 100 very massive stars, and, although we cannot provide a strict limit, it is hard to imagine that even more massive stars were formed in the GC. It does not seem likely that a significant non-disc population can be formed via scattering off such objects.

\begin{figure}
   \centerline{\epsfig{file=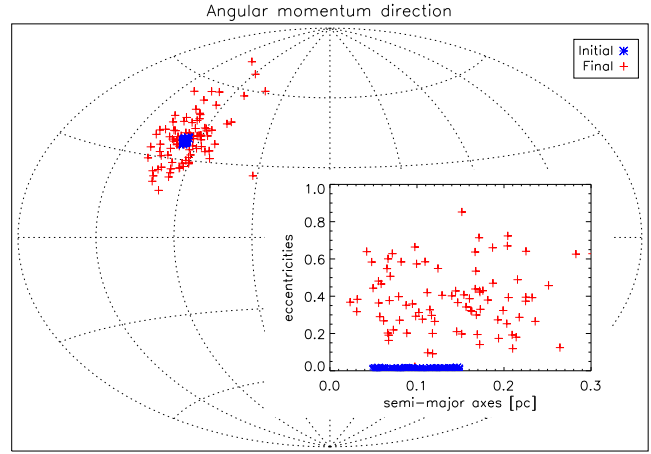,width=.49\textwidth}} 
   \caption{Aitoff projection of the angular momenta of the 25-$\msun$ stars in a simulation that also included 100 100-$\msun$ stars.  The angle definition is chosen such that the disc initially has the same orientation as the observed clockwise disc \citep[cf.,][Fig.~5]{Lu06}.  No star reaches  the bottom half of the plot, where they would be observed rotating counter-clockwise.  The inset shows the distribution of semi-major axes and eccentricities.}
   \label{fig:massive_angmom}
\end{figure}

To illustrate the failure of the model more clearly, Figure~\ref{fig:massive_angmom} shows the orbital parameters and angular momentum
direction of each 25~$\msun$ star from a run with 100 extra 100 $\msun$ stars.  Even in this case only a handful of stars seem to be outside
of the disc, and none of them appears to be rotating in the opposite direction.  The eccentricities, however, are
comparable to the observed values.

\subsection{Intermediate Mass Black Holes}
\label{sec:imbhs}

Finally, we investigated the possibility that scattering from a hypothetical population of 
intermediate mass black holes might heat the disc stars. \cite{PZwart06} argued that core 
collapse in GC star clusters could yield a significant population of such objects, with 
masses around 1000~$M_\odot$. Adopting their numbers, there could be $\sim$50 intermediate 
black holes within 10~pc of the GC, and some would likely have orbits that 
intrude upon the observed stellar disc.

To study the effect of these black holes, we set up simulations in which the stellar disc 
was initialised as described in Section~\ref{sec:singles}. We then 
added $N$ intermediate mass black holes, with orbits appropriate for an isotropic cluster 
with number density profile $n(r) \propto r^{-7/4}$ \citep{Bahcall76} (i.e. drawn from an energy 
distribution $f(E) \propto E^{1/4}$ with random directions). We consider a model with  
$N=5$ 1000-$M_\odot$ inside 1~pc (similar to the number expected from 
\citealt{PZwart06} analysis), together with additional cases that assume $N=10$ 500-$M_\odot$ and 
$N=20$ 500-$M_\odot$ intermediate mass black holes. 

\begin{figure}
   \centerline{\epsfig{file=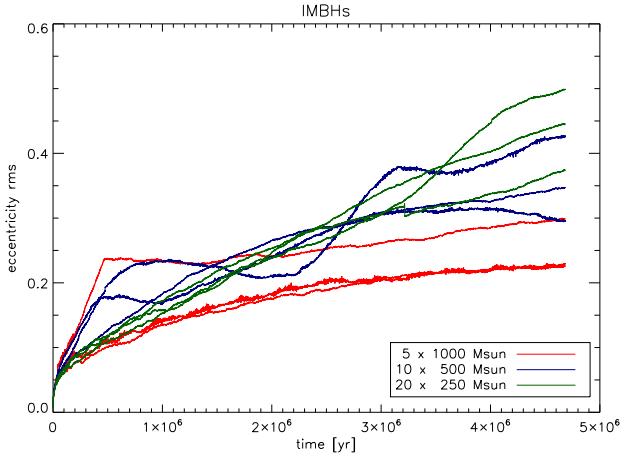,width=.49\textwidth}} 
   \caption{Eccentricity rms values from the nine different simulations of discs of single stars perturbed by IMBHs of different masses. The results are similar to those of the models with extra massive stars in the disc (Fig.~\ref{fig:massive}).}
   \label{fig:imbhs}
\end{figure}

The resulting $e_{\rm rms}$ curves are shown in Figure~\ref{fig:imbhs}. There is a lot of dispersion between the different 
realisations, since the small numbers of black holes involved mean that in some cases none (or 
very few) of the massive objects intersect the stellar disc. However, it is clear that the typical 
effect is rather similar to the case of massive stars discussed previously. Quite large eccentricities 
are readily produced, but the inclinations ($i_{\rm rms} \approx e_{\rm rms}/2$, not shown in the plot for clarity) remain too small to explain the observed distribution of 
stars in the Galactic Centre\footnote{Note that although \cite{Kozai62} resonance {\em would} allow for the 
generation of high inclinations, the rapid precession of stellar orbits due to the stellar cusp 
is likely to destroy this resonance.}.

\section{Conclusions}
In this paper we have studied the dynamical evolution of initially flat cold stellar discs consisting 
of the massive stars observed today in the GC. We find that these systems do not evolve to match the 
observed stellar population -- which includes significant numbers of stars that are not clearly 
identified with a well-defined disc plane -- even when scattering from plausible (and perhaps 
not so plausible) additional perturbers is included. It is clear that the most important observational 
constraints on theoretical models comes from measurements of the orbital inclinations. Although  
a flat circular disc does not become eccentric enough in 5~Myr to match the observations, it 
{\em is} possible to create quite eccentric systems either by starting from eccentric gas discs, 
or by perturbing circular discs with additional massive objects (short-lived stars, or 
intermediate mass black holes). However, in no cases that we have considered is it possible 
to generate enough very high inclination or counter-rotating stars within the available time. No physical process we have considered breaks the basic relation $e_{\rm rms} \approx 2 i_{\rm rms}$ between the mean eccentricity and 
inclination and this, coupled with the slow rate of heating at late times (typically $\propto t^{1/4}$), 
precludes formation of a significant non-disc population.

The simplest disc scenario for forming the sub-pc stars in the GC appeals to the fragmentation 
of a flat gaseous disc into a population of massive stars. Both the 
tendency of self-gravitating accretion discs at these radii to fragment, and the subsequent formation of 
a top-heavy mass function, receive at least qualified support from theory 
\citep{Shlosman89,Levin03,NCS07}. However, our results suggest that it is very unlikely 
that subsequent evolution of a stellar system formed in this manner would yield any 
substantial population of stars outside of the original disc plane or planes. Possible 
alternatives are that some or all of the GC stars migrated inward from 
larger radii (though this model has its own challenges, e.g., \citealt{Kim04,NS05});
that a non-spherical component in the background stellar population, or a second stellar disc \citep{NDCG06}, influences the dynamics of the main stellar disc in ways we have not considered;
or that in situ star formation occurs in a very complex geometry. In the latter case 
the present orbits of stars are expected to trace, to a first approximation, 
the angular momentum distribution of the star forming gas. Ideas that would 
yield dynamically hot initial conditions -- such as very strongly warped discs \citep{Nayakshin05w}
or clouds that collapse as they engulf the black hole \citep{Yusef08}, perhaps forming two coeval discs \citep{Genzel03a} -- have been proposed, though whether such scenarios can be analysed 
in the same framework as the simpler disc models is dubious.  

It is to be hoped that 
better observational data on the stellar orbits, which should confirm or refute 
the existence of additional clustering in the distribution of orbital vectors, 
will afford some insight into which models are viable.

\section*{Acknowledgements}

We thank Clovis Hopman, Mark Morris, and the anonymous referee for useful comments.
This research was supported by NASA under grants NNG04GL01G, NNX07AH08G and NNG05GI92G, and by the NSF under grant AST~0407040.  RDA acknowledges support from the Netherlands Organisation for Scientific Research (NWO) through VIDI grants 639.042.404 and 639.042.607.

\bibliography{biblio.bib}

\begin{thebibliography}{38}
\expandafter\ifx\csname natexlab\endcsname\relax\def\natexlab#1{#1}\fi

\bibitem[{Alexander} et~al.(2008){Alexander}, {Armitage}, {Cuadra} \&
  {Begelman}]{AACB08}
{Alexander} R.~D., {Armitage} P.~J., {Cuadra} J., {Begelman} M.~C., 2008, \apj,
  674, 927

\bibitem[{Alexander} et~al.(2007){Alexander}, {Begelman} \&
  {Armitage}]{Alexander07}
{Alexander} R.~D., {Begelman} M.~C., {Armitage} P.~J., 2007, \apj, 654, 907

\bibitem[{Bahcall} \& {Wolf}(1976)]{Bahcall76}
{Bahcall} J.~N., {Wolf} R.~A., 1976, \apj, 209, 214

\bibitem[{Beloborodov} et~al.(2006){Beloborodov}, {Levin}, {Eisenhauer}
  et~al.]{Beloborodov06}
{Beloborodov} A.~M., {Levin} Y., {Eisenhauer} F., et~al., 2006, \apj, 648, 405

\bibitem[{Berukoff} \& {Hansen}(2006)]{Berukoff06}
{Berukoff} S.~J., {Hansen} B.~M.~S., 2006, \apj, 650, 901

\bibitem[{Chambers}(1999)]{Chambers99}
{Chambers} J.~E., 1999, \mnras, 304, 793

\bibitem[{Gammie}(2001)]{Gammie01}
{Gammie} C.~F., 2001, \apj, 553, 174

\bibitem[{Genzel} et~al.(2003){Genzel}, {Sch{\" o}del}, {Ott}
  et~al.]{Genzel03a}
{Genzel} R., {Sch{\" o}del} R., {Ott} T., et~al., 2003, \apj, 594, 812

\bibitem[{Gerhard}(2001)]{Gerhard01}
{Gerhard} O., 2001, \apj, 546, L39

\bibitem[{Goodman}(2003)]{Goodman03}
{Goodman} J., 2003, \mnras, 339, 937

\bibitem[{G{\"u}rkan} \& {Rasio}(2005)]{Gurkan05}
{G{\"u}rkan} M.~A., {Rasio} F.~A., 2005, \apj, 628, 236

\bibitem[{Heggie}(1975)]{Heggie75}
{Heggie} D.~C., 1975, \mnras, 173, 729

\bibitem[{Hopman} \& {Alexander}(2006)]{Hopman06}
{Hopman} C., {Alexander} T., 2006, \apj, 645, 1152

\bibitem[{Kim} et~al.(2004){Kim}, {Figer} \& {Morris}]{Kim04}
{Kim} S.~S., {Figer} D.~F., {Morris} M., 2004, \apj, 607, L123

\bibitem[{Kim} \& {Morris}(2003)]{Kim03}
{Kim} S.~S., {Morris} M., 2003, \apj, 597, 312

\bibitem[{Kolykhalov} \& {Sunyaev}(1980)]{Kolykhalov80}
{Kolykhalov} P.~I., {Sunyaev} R.~A., 1980, Soviet Astron. Lett., 6, 357

\bibitem[{Kozai}(1962)]{Kozai62}
{Kozai} Y., 1962, \aj, 67, 591

\bibitem[{Levin}(2007)]{Levin07}
{Levin} Y., 2007, \mnras, 374, 515

\bibitem[{Levin} \& {Beloborodov}(2003)]{Levin03}
{Levin} Y., {Beloborodov} A.~M., 2003, \apj, 590, L33

\bibitem[{Lissauer}(1993)]{Lissauer93}
{Lissauer} J.~J., 1993, \araa, 31, 129

\bibitem[{Lu} et~al.(2006){Lu}, {Ghez}, {Hornstein} et~al.]{Lu06}
{Lu} J.~R., {Ghez} A.~M., {Hornstein} S.~D., et~al., 2006, Journal of Physics
  Conference Series, 54, 279

\bibitem[{Martins} et~al.(2006){Martins}, {Trippe}, {Paumard}
  et~al.]{Martins06}
{Martins} F., {Trippe} S., {Paumard} T., et~al., 2006, \apjl, 649, L103

\bibitem[{Milosavljevi{\' c}} \& {Loeb}(2004)]{Milosavljevic04}
{Milosavljevi{\' c}} M., {Loeb} A., 2004, \apj, 604, L45

\bibitem[{Nayakshin}(2005)]{Nayakshin05w}
{Nayakshin} S., 2005, \mnras, 359, 545

\bibitem[{Nayakshin} \& {Cuadra}(2005)]{NC05}
{Nayakshin} S., {Cuadra} J., 2005, \aap, 437, 437

\bibitem[{Nayakshin} et~al.(2007){Nayakshin}, {Cuadra} \& {Springel}]{NCS07}
{Nayakshin} S., {Cuadra} J., {Springel} V., 2007, \mnras, 379, 21

\bibitem[{Nayakshin} et~al.(2006){Nayakshin}, {Dehnen}, {Cuadra} \&
  {Genzel}]{NDCG06}
{Nayakshin} S., {Dehnen} W., {Cuadra} J., {Genzel} R., 2006, \mnras, 366, 1410

\bibitem[{Nayakshin} \& {Sunyaev}(2005)]{NS05}
{Nayakshin} S., {Sunyaev} R., 2005, \mnras, 364, L23

\bibitem[{Paczy\'nski}(1978)]{Paczynski78}
{Paczy\'nski} B., 1978, Acta Astron., 28, 91

\bibitem[{Paumard} et~al.(2006){Paumard}, {Genzel}, {Martins}
  et~al.]{Paumard06}
{Paumard} T., {Genzel} R., {Martins} F., et~al., 2006, \apj, 643, 1011

\bibitem[{Peeples} et~al.(2007){Peeples}, {Bonanos}, {DePoy} et~al.]{Peeples07}
{Peeples} M.~S., {Bonanos} A.~Z., {DePoy} D.~L., et~al., 2007, \apjl, 654, L61

\bibitem[{Perets} et~al.(2008){Perets}, {Kupi} \& {Alexander}]{Perets07}
{Perets} H.~B., {Kupi} G., {Alexander} T., 2008, in { Dynamical Evolution of
  Dense Stellar Systems\/}, vol. 246 of { IAU Symposium\/},  275--276

\bibitem[{Portegies Zwart} et~al.(2006){Portegies Zwart}, {Baumgardt},
  {McMillan}, {Makino}, {Hut} \& {Ebisuzaki}]{PZwart06}
{Portegies Zwart} S.~F., {Baumgardt} H., {McMillan} S.~L.~W., {Makino} J.,
  {Hut} P., {Ebisuzaki} T., 2006, \apj, 641, 319

\bibitem[{Rafelski} et~al.(2007){Rafelski}, {Ghez}, {Hornstein}, {Lu} \&
  {Morris}]{Rafelski07}
{Rafelski} M., {Ghez} A.~M., {Hornstein} S.~D., {Lu} J.~R., {Morris} M., 2007,
  \apj, 659, 1241

\bibitem[{Rauch} \& {Tremaine}(1996)]{Rauch96}
{Rauch} K.~P., {Tremaine} S., 1996, New Astronomy, 1, 149

\bibitem[{Rice} et~al.(2003){Rice}, {Armitage}, {Bonnell}, {Bate}, {Jeffers} \&
  {Vine}]{Rice03}
{Rice} W.~K.~M., {Armitage} P.~J., {Bonnell} I.~A., {Bate} M.~R., {Jeffers}
  S.~V., {Vine} S.~G., 2003, \mnras, 346, L36

\bibitem[{Shlosman} \& {Begelman}(1989)]{Shlosman89}
{Shlosman} I., {Begelman} M.~C., 1989, \apj, 341, 685

\bibitem[{Yusef-Zadeh} \& {Wardle}(2008)]{Yusef08}
{Yusef-Zadeh} F., {Wardle} M., 2008, in { Massive Star Formation: Observations
  Confront Theory\/}, vol. 387 of { Astronomical Society of the Pacific
  Conference Series\/},  361--+

\end{thebibliography}
\bibliographystyle{mnras}

\label{lastpage}

\end{document}